\documentclass[aps,prd,groupedaddress]{revtex4-2}

\usepackage{graphicx}
\usepackage{amsmath}
\usepackage{verbatim}
\usepackage{array}
\usepackage{booktabs}

\begin{document}




\title{ Determination of the resonant parameters of $\chi_{c0}(3915)$ with global fit}

\author{Chunhua Li$^{1,2}$\thanks{chunhua@lnnu.edu.cn}, Xijun Wang$^{1,2}$, Chen Wu$^{1,2}$\\
$^1$Department of physics and electronic technology,\\
$^2$Center for Theoretical and Experimental High Energy Physics,\\
Liaoning Normal University\\
Dalian, 116029, P.R.China\\
}

\begin{abstract}
The particle $\chi_{c0}(3915)$ was first observed by the Belle experiment in the $\omega J/\psi$ invariant mass spectrum in the process 
$B\to K\omega J/\psi$, and subsequently confirmed by the BaBar experiment.
The two experiments reported the resonant parameters of this particle in both of the processes
$\gamma\gamma\to\omega J/\psi$ and $B\to K\omega J/\psi$ by assuming $\chi_{c0}(3915)$ as an S-wave Breit-Wigner resonance.
We perform a global fit to the distributions 
of invariant mass of $\omega J/\psi$ 
measured by the Belle and BaBar experiments, and incorporate the measurement by the LHCb experiment additionally to extract 
the mass and width of $\chi_{c0}(3915)$. 
We obtain $M=3920.9\pm0.9$~MeV/$c^2$ and $\Gamma=18.2\pm2.4$~MeV, which are consistent with the values on PDG within one standard deviation but with improved precision.
\end{abstract}


\maketitle



\section{Introduction}
Since the first charmonium-like state $X(3872)$ was discovered by 
Belle experiment via the process $e^+e^-\to\gamma_{ISR}\pi^+\pi^- J/\psi$, 
a new era of the study of charmonium-like states was triggered~\cite{X1}.
A series of mesons comprised of charmed and anticharmed quark pairs,
such as the Y(4260), $X(3915)$, and Zc(3900), were subsequently discovered, 
and most 
of them were confirmed by different experiments~\cite{Y1,Y2,Y3,Z1,Z2,Belle-gg,BaBar-gg,
Belle-B,BaBar-B1,BaBar-B2}. 
The understanding of the nature of these particles has turned out to be quite
a challenge.
Many of them have properties that are quite different from the conventional 
charmonium, e.g. a low open-charm decay rate or the absence of hadronic 
transitions to other charmonium states.
To explain these anomalous features, many models have been proposed by theorists, 
including charmonium molecule mixed states~\cite{M1,M2}, $c\bar{c}g$ hybrid states~\cite{H1}, 
and tetra-quarks~\cite{T1}.
In addition, the production rates for some of these charmonium-like states are
quite low in most experiments compared to that of the conventional charmonium states, which limits the measurement precision of  
the resonant parameters and the determination of the
corresponding quantum numbers. This makes 
the interpretation of these particles difficult.  
Particle Data Group (PDG)~\cite{pdg} renamed the $X(3915)$ and $X(3872)$  
as $\chi_{c0}(3915)$ and $\chi_{c1}(3872)$, respectively, according to 
their spin-parities. We just follow the PDG 
naming convention in the article.

As a member of charmonium-like family of states, the
$\chi_{c0}(3915)$ was first observed by the Belle experiment 
in the process $B\to K J/\psi\omega$ 
in a data sample containing $275\times10^6$ $B\bar{B}$ pairs~\cite{Belle-B}. 
The mass and width were determined to be 
$3943\pm11\pm13$~MeV/$c^2$ and $87\pm22\pm26$~MeV 
with the assumption that the 
$\chi_{c0}(3915)$ is an S-wave Breit-Wigner (BW) resonance. The particle 
was confirmed by BaBar experiment in the same decay mode
with a $383\times10^6$ $B\bar{B}$ event data sample~\cite{BaBar-B1}, 
their reported mass and width are $3914.6^{+3.8}_{-3.4}\pm2.0$~MeV/$c^2$
and $34^{+12}_{-8}\pm5$~MeV. 
BaBar experiment subsequently updated their measurements 
with a lager data sample of $467\times10^{6}$ $B\bar{B}$ events
and looser $M(\pi^+\pi^-\pi^0)$ requirement that revealed a 
X(3872) signal as well~\cite{BaBar-B2}; 
the updated mass and width are $3919.1^{+3.8}_{-3.4}\pm2.0$~MeV/$c^2$
and $31^{+10}_{-8}\pm5$~MeV.
In addition, the $\chi_{c0}(3915)$ was also observed in the 
two-photon collision process 
$\gamma\gamma\to\omega J/\psi$ by both the BaBar and Belle experiments~\cite{Belle-gg,BaBar-gg}.
Their measured masses and widths are listed in Table~\ref{ta1}.
BaBar performed a spin-parity measurement in their analysis,  obtaining the quantum number to 
be $J^P=0^+$ and identifying the $\chi_{c0}(3915)$ as the $\chi_{c0}(2P)$ 
charmonium state.
However, this assignment was disputed because of the large rate for
the $\chi_{c0}(3915)\to\omega J/\psi$ decay and the absence 
of $\chi_{c0}(3915)\to D\bar{D}$ decays~\cite{guofk,steve}.
Moreover, the mass difference between the 
$\chi_{c2}(2P)$ and $\chi_{c0}(3915)$ is only about 10 MeV, which 
is too small for the fine splitting of $P$-wave charmonia~\cite{guofk}.
In 2020, LHCb experiment made an amplitude analysis of the $B^+\to D^+D^-K^+$ decay~\cite{lhcb}, and reported that
a spin-0 resonance is needed to describe the data well. They determined its mass and width to be 
$2923.8\pm1.5\pm0.4$~MeV/$c^2$ and $17.4\pm5.1\pm0.8$~MeV, respectively.

\begin{figure*}[h]
\includegraphics[width=0.45\textwidth]{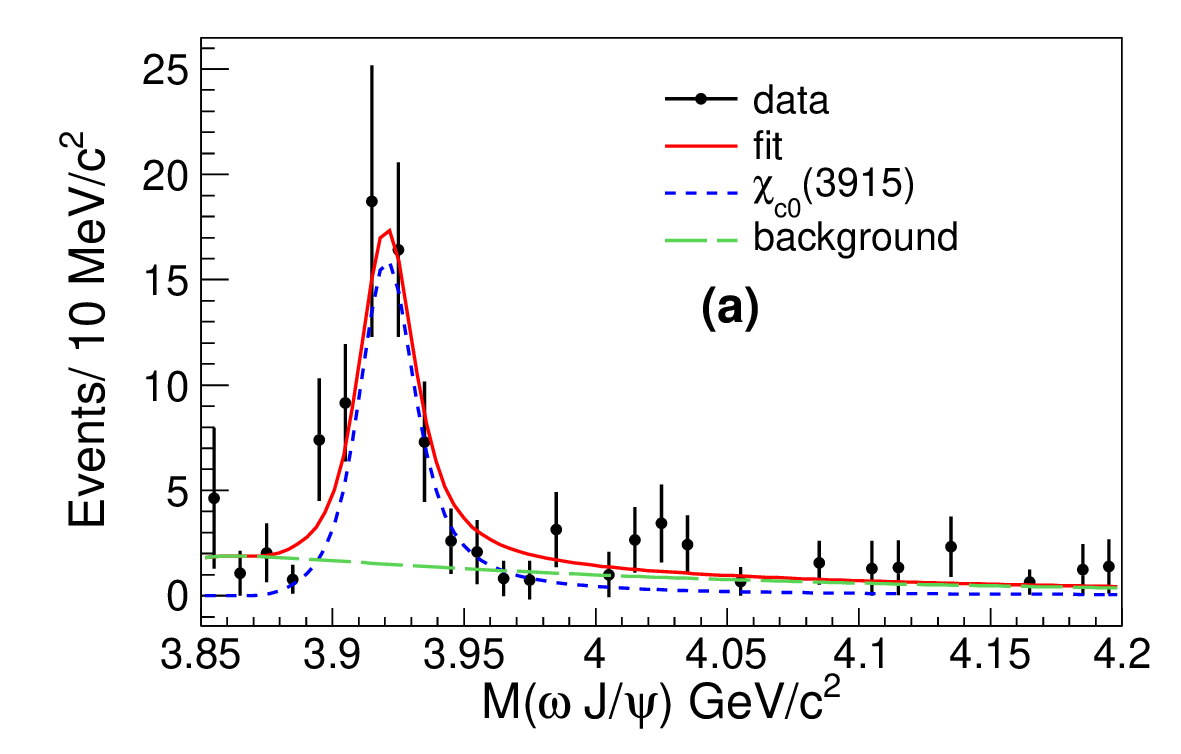}
\includegraphics[width=0.45\textwidth]{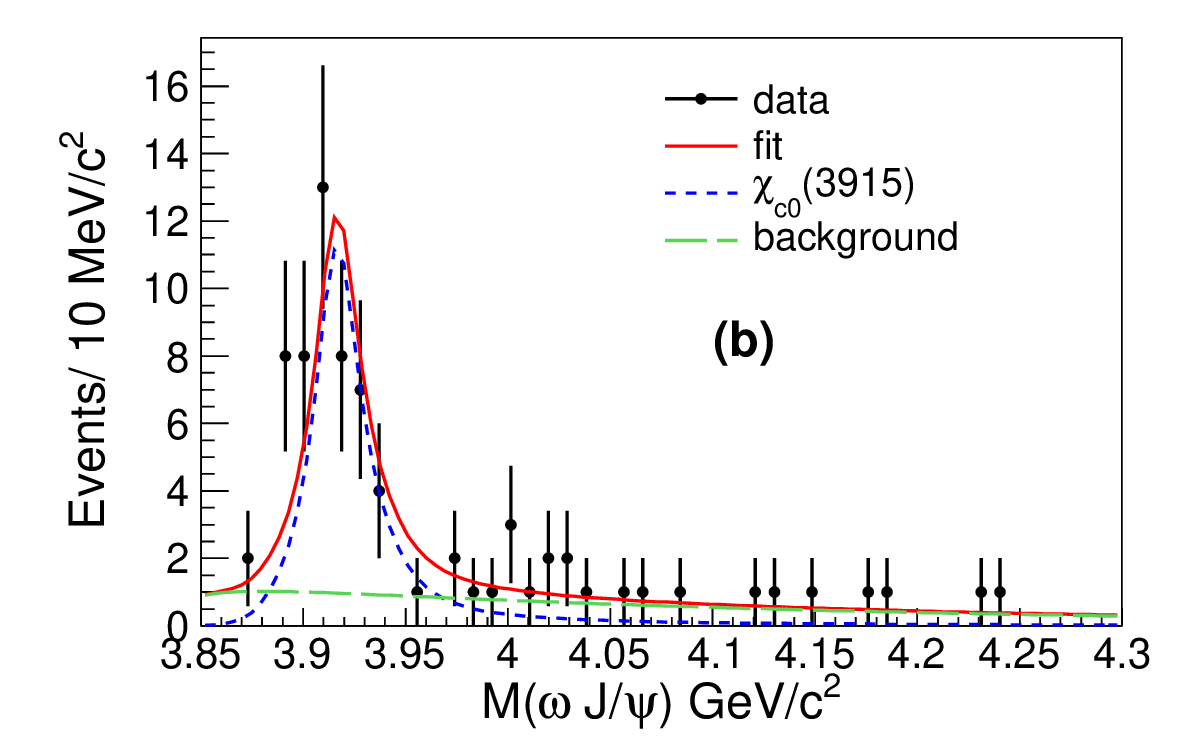}
\includegraphics[width=0.45\textwidth]{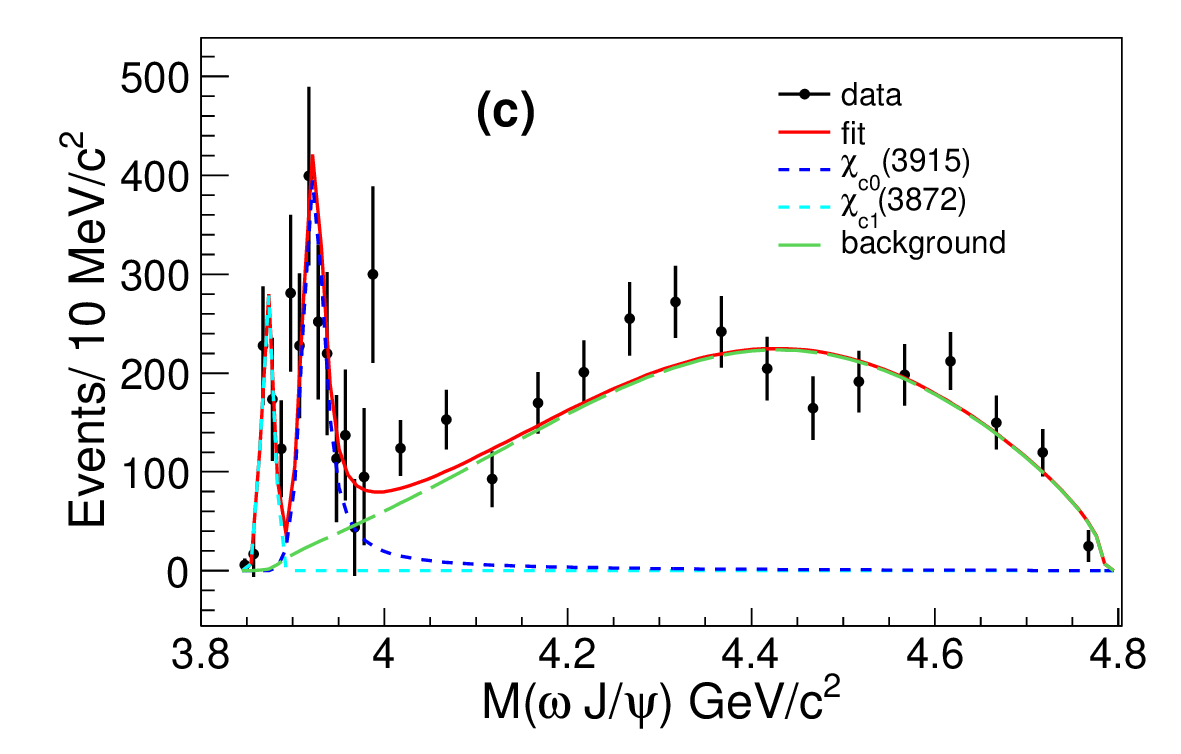}
\includegraphics[width=0.45\textwidth]{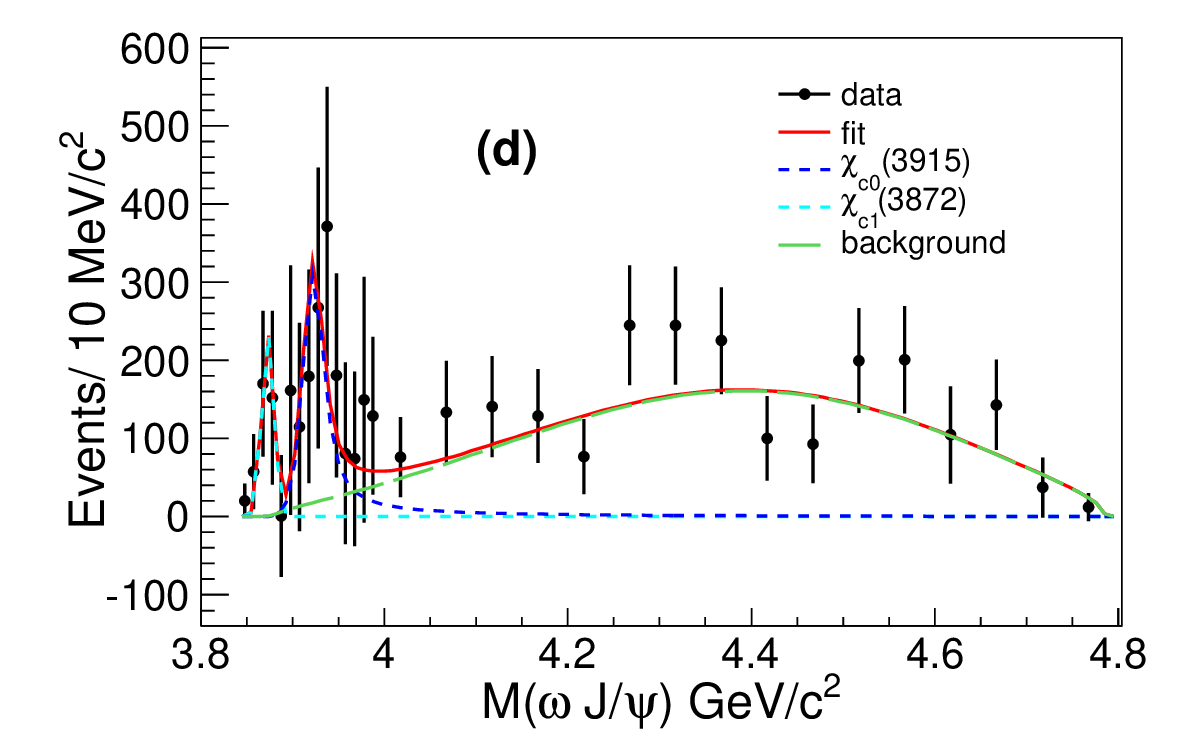}
\includegraphics[width=0.45\textwidth]{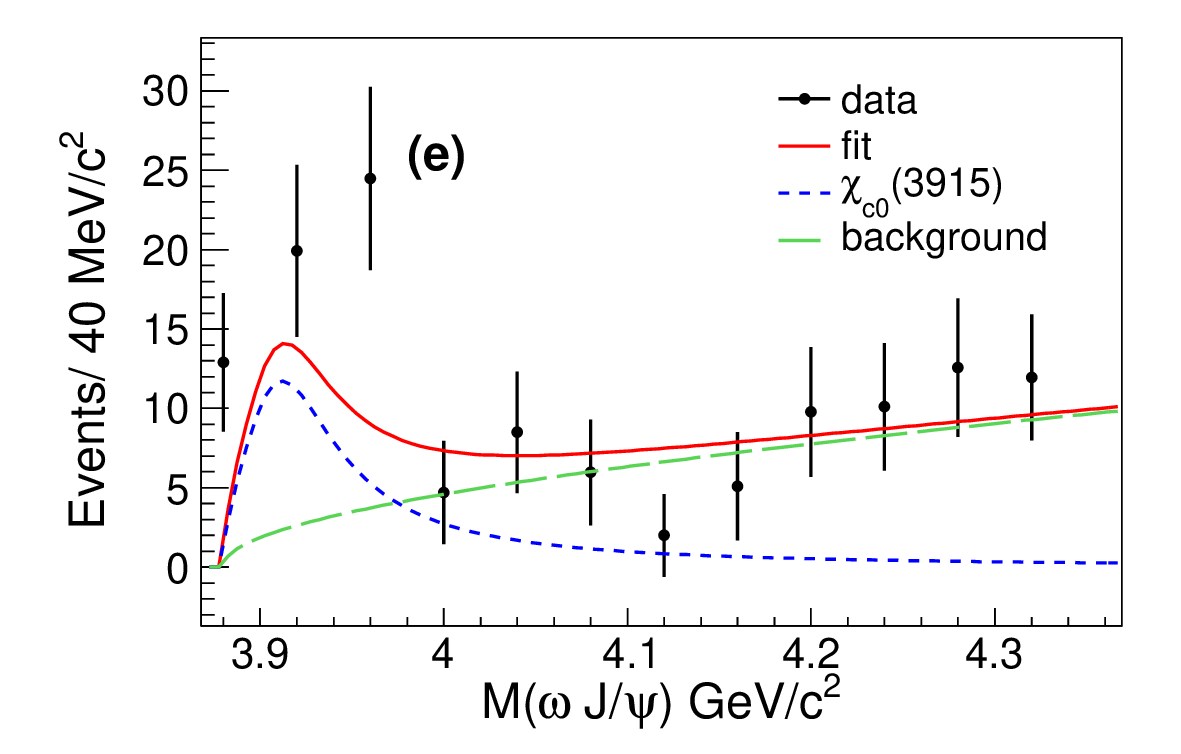}
\caption{
Fit to the distributions of the $\omega J/\psi$ invariant mass obtained in the processes
(a) $\gamma\gamma\to \omega J/\psi$ by BaBar, 
(b) $\gamma\gamma\to \omega J/\psi$ by Belle, 
(c) $B^0\to \omega J/\psi K^0$ by BaBar, (d) $B^+\to \omega J/\psi K^+$
by BaBar, and (e) $B\to \omega J/\psi K$ by Belle.
The black dots with error bars are the data mentioned in the main text.
The red solid curves are the fit results, 
the cyan/blue dashed curves indicate the $\chi_{c1}(3872)$/$\chi_{c0}(3915)$ signals, and 
the green dashed curves show the backgrounds.
\label{fig1}}
\end{figure*}

In this article, we perform a simultaneous $\chi^2$ fit to 
the distributions of invariant mass of $\omega J/\psi$ in the 
processes $\gamma\gamma\to \omega J/\psi$ measured by BaBar [denoted as (a)], 
$\gamma\gamma\to \omega J/\psi$ by Belle [(b)], 
$B^0\to \omega J/\psi K^0$ by BaBar [(c)], $B^+\to \omega J/\psi K^+$[(d)]
by BaBar, and $B\to \omega J/\psi K$ by Belle [(e)] to extract 
the mass and width of $\chi_{c0}(3915)$. The distributions of $M(\omega J/\psi)$ for these processes
are shown in Fig.~\ref{fig1}. Furthermore, LHCb's results are taken into account as an additional 
constraint in the $\chi^2$ calculation.
Compared to the values on PDG, which also gives the $\chi_{c0}(3915)$ mass and width 
by combining the measurements from these experiments, we use more detailed information of the $\omega J/\psi$
invariant mass spectrum, which is expected to provide a result with higher precision.

\section{Simultaneous $\chi^2$ fit}
In the simultaneous $\chi^2$ fit, the functions used to fit the $M(\omega J/\psi)$ spectra
in Fig.~\ref{fig1} are similar to those applied in their corresponding publications.
The fit function is comprised of signal and background components.
The $\chi_{c0}(3915)$ signal shape is described by an S-wave BW function
convoluted with the detector resolutions which are also from 
the publications. The BW is 
$\frac{\Gamma(p^*/p_{0})}{(m^2-M^2)^2+(M\Gamma(p^*/p_{0}))^2}$,
where $M$ is the peak mass, $p^*$ is the momentum of $J/\psi$ momentum
in the rest frame of a $\omega J/\psi$ system, and $p_0=p^*$ when $m=M$~\cite{x3915BW}.
$M$ and $\Gamma$ are common parameters for the five measurements that are 
allowed to float in the fit.
The contributions of the $\chi_{c1}(3872)$ are needed for the distributions (c) and (d) as shown 
in Fig.~\ref{fig1}.
The signal shape of the $\chi_{c1}(3872)$ is described with a Gaussian function with a fixed deviation of 6.7~MeV/$c^2$ but with a free mean value.
The detector resolution of $\gamma\gamma\to\omega J/\psi$ for Belle 
is described by a double Gaussian function, one Gaussian has the mean and deviation values 
of 4.5 MeV and 0 MeV, respectively, with a coefficient of 0.59,
and the other one has a mean and deviation of 16 MeV and 
-4.0 MeV with a cofficient of 0.41~\cite{Belle-gg}. 
Conversely, the resolution for the same process by BaBar 
is described by a single Gaussian with a deviation of $5.7$~MeV and a mean value of zero.
Similarly, the resolution for the measurement $B\to K\omega J/\psi$ 
by BaBar (Belle) is described by a single Gaussian function 
with a deviation of 6.7~MeV (6~MeV) and a mean value of zero.
All these resolution details are from the corresponding publications.
The non-resonance background shapes for the measurement (a) and (b) are
described as $p^*(m){\rm exp}(-\delta p^*(m))$, where $\delta$ 
is a parameter that is allowed to float 
in the fit, and $m=m(J/\psi\omega)$~\cite{BaBar-gg}.
The background shapes of the measurement (c) and (d) are described 
with a Gaussian function with parameters that float in the fit~\cite{BaBar-B2}.
The shape of the background in the measurement (e) is described with the threshold function 
of the form $p^*(m)$, which is the $J/\psi$ momentum
in the rest frame of a $\omega J/\psi$ system.

The $\chi^2$ is defined as 
\begin{equation}
\chi^2 = \sum_{i=1}^{5}\sum_{j=1}^{N_i} (\frac{x_{ij}-\mu_{ij}}{\sigma_{ij}})^2+(\frac{M_{h}-M}{\sigma_{M_{h}}})^2+(\frac{\Gamma_h-\Gamma}{\sigma_{\Gamma_{h}}})^2
\end{equation}
where $i$ takes values from 1 to 5, corresponding 
to the measurements from (a) to (e), 
$x_{ij}$ and $\sigma_{ij}$ are the observed signals and 
corresponding errors in each bin as shown in Fig.~\ref{fig1}, 
$\mu_{ij}$ is the expected value in each bin 
calcluated with the fitting shape mentioned above, and
$N_i$ is the number of bins in each measurement.
The numbers of events in many bins for the measurements (a), (b), and (e) are 
very small,
so we reset the bin width to 
ensure that there is at least nine events in each bin to 
make a meaningful calculation of their contribution to 
the $\chi^2$ value. The numbers of bins 
in the measurement (a) and (b) are $N_1$=6 and $N_2$=8.
For the measurements (c) and (d), the bin width in the mass region of 3.8425 to 3.9925~GeV/$c^2$ 
is 10~MeV/$c^2$, and 50~MeV/$c^2$ in the region beyond 3.9925 GeV/$c^2$ as shown in
Fig.~\ref{fig1}. The numbers of bins $N_3$ and $N_4$ are both 31. For the measurement (e), $N_5$=9.
The last two components of the $\chi^2$ formula, $\frac{M_{h}-M}{\sigma_{M_{h}}}$ and $\frac{\Gamma_{h}-\Gamma}{\sigma_{\Gamma_{h}}}$,
are from the measurements of the LHCb experiment~\cite{lhcb}, where $M_h$ and $\Gamma_h$ are 
the measured mass and width, and $\sigma_{M_h}$ and $\sigma_{\Gamma_h}$ the 
corresponding statistical uncertainties. 
 
By minimizing the $\chi^2$ with {\sc minuit}~\cite{minui}, 
we obtain the fit result 
with $M=3920.9\pm0.8$~MeV/$c^2$ and $\Gamma=18.2\pm2.2$~MeV. 
The goodness of fit is $\chi^2/ndf= 89.1/66$ where $ndf$ 
is the number of degrees of freedom in the fit. The fit results are also shown 
in Fig.~\ref{fig1}. 

The total systematic uncertainty of the mass and width, $\sigma_{\rm sum}$, is obtained with 
the formula $\frac{1}{\sigma_{\rm sum}^2}=\sum_{i=1}^{5}\frac{1}{\sigma_{i}^2}$, where the $i$ has a value from 1 to 5 
corresponding to each measurement listed in Table~\ref{ta1}, and $\sigma_{i}$ is the systematic uncertainty of the $i$-th measurement.
Combining with the statistical uncertainties from our fit, we finally get the results of 
$M=3920.9\pm0.9$~MeV/$c^2$ and $\Gamma=18.2\pm2.4$~MeV.

\begin{table*}[htbp]
\caption{Mass ($M$) and width ($\Gamma$) of the $\chi_{c0}(3915)$ measured by different experiments, where the first uncertainty is the statistical 
and the second is systematic. The results from PDG and this work are listed in the last two rows.}
\begin{center}
\begin{tabular}{@{}llcc@{}} \toprule
Experiment &Production& $M$ (MeV/$c^2$)& $\Gamma$ (MeV)  \\
\hline
Belle~\cite{Belle-B}&$B\to\omega J/\psi K$& $3943\pm11\pm13$& $87\pm22\pm26$\\
BaBar~\cite{BaBar-B2}&$B\to\omega J/\psi K$ &$3919.1^{+3.8}_{-3.4}\pm2.0$ &$31^{+10}_{-8}\pm5$ \\
Belle~\cite{Belle-gg}&$\gamma\gamma\to\omega J/\psi$ & $3915\pm3\pm2$ & $13\pm6\pm3$ \\
BaBar~\cite{BaBar-gg}&$\gamma\gamma\to\omega J/\psi$ & $3919.4\pm2.2\pm1.6$& $17\pm10\pm3$\\
LHCb~\cite{lhcb}&$B^+\to D^+D^-K^+$ &$3923.8\pm1.5\pm0.4$&$17.4\pm5.1\pm0.8$\\
\hline
PDG~\cite{pdg} &&$3921.7\pm1.8$&$18.8\pm3.5$\\
This work&&$3920.9\pm0.9$&$18.2\pm2.4$\\
\hline
\end{tabular}\label{ta1} 
\end{center}
\end{table*}

\section{Summary}

In summary, we determine the resonant parameters of $\chi_{c0}(3915)$ 
by simultaneously fitting the measurements 
provided by the Belle, BaBar, and LHCb experiments. 
The mass and width are determined 
to be $M=3920.9\pm0.9$~MeV/$c^2$ and $\Gamma=18.2\pm2.4$~MeV, respectively. 
which are consistent with
the average values in PDG as listed in Table~\ref{ta1} within 
one standard deviation but with improved precision.

\section*{Acknowledgments}
This work is supported by 
National Natural Science Foundation of China (NSFC) 
under contract No. 11805090.

\end{document}